\documentclass[aps,11pt,prc,preprint,superscriptaddress,nofootinbib]{revtex4}
\usepackage[usenames]{color}
\usepackage{graphicx}
\usepackage{amsmath}
\usepackage{amsfonts}
\usepackage{amssymb}
\usepackage{mathrsfs}
\usepackage{bm}
\usepackage{verbatim}

\usepackage{ulem}

\newcommand{\beq}{\begin{equation}}
\newcommand{\eeq}{\end{equation}}
\newcommand{\bea}{\begin{eqnarray}}
\newcommand{\eea}{\end{eqnarray}}

\graphicspath{{../}}

\begin{document}

\title{Toward a systematic strategy for defining power counting in the construction of the energy density functional theory}
\author{C.-J.~Yang}
\affiliation{Institut de Physique Nucl\'eaire, CNRS/IN2P3, Universit\'e Paris-Sud, Universit\'e Paris-Saclay, F-91406, Orsay, France}
\author{M. Grasso}
\affiliation{Institut de Physique Nucl\'eaire, CNRS/IN2P3, Universit\'e Paris-Sud, Universit\'e Paris-Saclay, F-91406, Orsay, France}
\author{D. Lacroix}
\affiliation{Institut de Physique Nucl\'eaire, CNRS/IN2P3, Universit\'e Paris-Sud, Universit\'e Paris-Saclay, F-91406, Orsay, France}
\email{yangjerry@ipno.in2p3.fr, grasso@ipno.in2p3.fr,lacroix@ipno.in2p3.fr}
\date{\today }

\begin{abstract}
We propose a new scheme for constructing an effective--field--theory--based interaction to be used in the energy--density--functional (EDF) theory with specific assumptions for defining a power counting.
This procedure is developed through the 
evaluation of the equation of state (EOS) of symmetric and pure neutron matter going beyond the mean--field scheme and using a functional defined up to next--to--leading order (NLO), that we will call NLO EDF. 
A Skyrme--like
interaction is constructed based on the condition of renormalizibility and on a
power counting on $k_F/\Lambda_{hi}$, where $k_F$ is the Fermi momentum and $\Lambda_{hi}$ is the breakdown scale of our expansion. To absorb the divergences present in beyond mean--field diagrams, counter interactions are introduced for the NLO EDF and determined through renormalization conditions. 
In particular, three scenarios are explored and all of them lead to satisfactory results. 
These counter interactions contain also parameters which do not contribute to the EOS of matter and may eventually be determined through future adjustments to properties of some selected finite nuclei. Our work serves as a simple starting point for constructing a well--defined power counting 
within the EDF framework. 
\end{abstract}

\maketitle

\vspace{10mm}

\section{Introduction}

\bigskip The nuclear many--body problem has been extensively investigated since 
several decades. One of the challenges, at a very fundamental level, is the 
development of the
nucleon-nucleon (NN) interaction. Several versions of phenomenological and recently 
developed chiral effective--field--theory (EFT) potentials have been applied
to nuclear matter calculations through various ab-inito methods \cite%
{qmc1,qmc2,qmc3,db-nm,qmc5,lattice-eft,resum-k,qmcp1,qmcp2,qmcp3,qmcp4,qmcp5,qmce1,qmce2,qmce3,qmce4,qmce5,qmce6,qmce7,qmce8,qmce9,rev}%
. However, full convergence with respect to either the
method or the version of the potential is not yet achieved. Moreover, although 
relevant progress was recently made to extend the area of applicability of ab--initio methods 
\cite{Bog14,Her14,Sig15,Geb16,Jan16,Str16,Str17,Tic17}, it is not clear 
whether such methods can indeed be applied in future to the full 
nuclear chart, up to heavy nuclei. On the other hand, 
EDF theories have been adopted in nuclear many--body calculations for several 
decades with reasonable results \cite{bender}. In this approach, one does not start
from the bare interaction between nucleons and assumes the validity of a 
mean--field (or beyond--mean--field) picture, in most cases constructed using  
effective phenomenological 
interactions.
The Skyrme interaction \cite{skyrme,vauth} is 
one of the most popular choices adopted in EDF. It consists of series of
zero-range terms expanded in powers of momentum, which have an identical
form (except for the density--dependent term) as the contact interactions present in pionless EFT \cite{pionless,pastore}. 
The success of Skyrme--based 
calculations in the EDF
framework suggests that an EFT--like expansion based on a series of contact--type
terms may exist, and results obtained at the 
mean--field level may be chosen to represent
the leading--order (LO) contribution in such an expansion for EDF.\footnote{Additional indication is provided in Ref. \cite{denis17}, where it is shown that the magnitude of various versions of Skyrme coefficients can be recovered by an expansion based on the unitarity limit.} 

To further
explore along this direction, higher--order corrections need to be included. 
For example, in
Refs. \cite{mog,ym,kaiser,mog1}, the second--order contribution to the EOS of nuclear matter is derived 
analytically for Skyrme--type interactions. It is shown that, with
the inclusion of a density--dependent term, a reasonable 
EOS can be obtained for matter up to second order at various isospin asymmetries after the divergence is subtracted in various ways.
Furthermore, Ref. \cite{bira} shows that requirements based on renormalizability
restrict the Skyrme interaction to have certain forms. In particular, only the $t_{0}$ or $t_{0}-t_{3}$ Skyrme--type interactions with some specific powers of the density $\alpha$ are allowed for the second--order EOS to be renormalizable. In practice, only the latter interaction ($t_{0}-t_{3}$ model) could provide an acceptable second--order EOS for symmetric matter. Note that, except for the finite part, contributions from second--order diagrams are regularization-scheme-dependent. Whereas pionless EFT can be easily 
applied to vacuum or to dilute neutron matter and results become regularization-scheme independent after the renormalization is performed (for example, the free parameters can be matched to the effective--range expansion in the case of dilute neutron matter \cite{lee,hammer,hammerlucas,Fur12,pug,fur,bishop}), the renormalization/regularization process is more involved in the case of nuclear matter at larger densities. In particular, it is shown that if one considers symmetric nuclear matter at densities 
around the equilibrium point and starts with a Skyrme--like interaction, second--order results
 depend on the regularization procedure quite strongly \cite{bira}. In Ref. \cite{bira}, the conventional definition of effective mass at the mean--field level was adopted, and no additional contact interactions were added (no counter terms).  
 
In this work, we do not use a mean--field effective mass, we add counter terms by defining NLO effective interactions, and, as a consequence, we do not need to constrain the values of the density dependence $\alpha$. Starting from a $t_0-t_3$ model and the related contributions up to NLO, and guided by 
renormalizibility and renormalization--group (RG) analysis, we explore three types of possible counter terms and
develop the EOS for symmetric and neutron matter up to NLO in EDF.

The structure of the present work is as follows. In section II, we describe
the theoretical framework of our approach and report the LO results. In section III, we apply our
method to develop a new Skyrme-like EFT interaction up to NLO 
and discuss the results. We summarize our findings in section IV.

\section{Theoretical framework}

\subsection{General considerations}

We first clarify the notation that we use in this work for LO and NLO.

\begin{figure}[h]
\includegraphics[width=6cm]{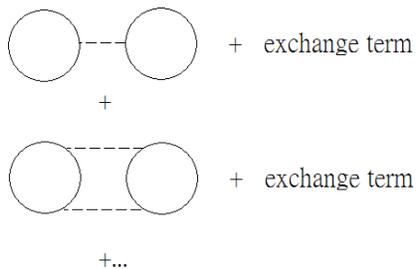}
\caption{Perturbative expansion of the ground--state energy in a uniform system. The diagramatic analysis of many--body perturbation theory is for instance illustrated in Ref. \cite{FW}.}
\label{j3}
\end{figure}
\begin{figure}[h]
\includegraphics[width=6cm]{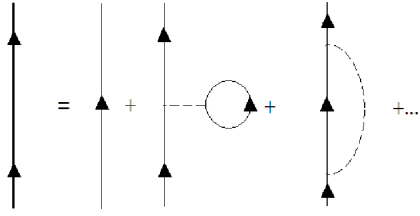}
\caption{Perturbative expansion of the exact Green's function. We refer to Ref. \cite{FW} for details on the 
diagrammatic representation of the many--body perturbation theory.}
\label{j1}
\end{figure}
Starting from a given NN interaction, the EOS of nuclear matter can be evaluated by summing 
the diagrams of the perturbative expansion of the energy shown in Fig.\ref{j3}. 
The 
diagrams to obtain the dressed propagator $G$ (the exact Green's function) are shown in Fig. \ref{j1}. 
Figures \ref{j3} and \ref{j1} represent the usual many--body perturbative expansion for the energy 
and the Green's function, respectively. In particular, the upper part of Fig. \ref{j3} describes the LO (first order or mean field) and the lower part the NLO (second order) of such a many--body expansion for the evaluation of the energy.

On the other hand,
for very dilute neutron matter (densities $\rho < 10^{-6}$ fm$^{-3}$), one
can perform a perturbative calculation based on the effective--range
expansion of the interaction, where higher loops are suppressed by higher
powers of $ak_{F}$, $a$ being the neutron--neutron s-wave scattering length, and can obtain physical
observables at very low densities \cite{lee,hammer,hammerlucas,Fur12,pug,fur,bishop}. However,
most of nuclear systems of interest have a density $\rho $ much higher than the dilute
limit. For example, typical densities in nuclear matter (of interest for finite nuclei) cover
the range  $%
\rho =0\sim 0.3$ fm$^{-3}$. To perform calculations at such densities one needs
to use other procedures. A density--dependent neutron-neutron scattering length was for instance 
adopted in Ref. \cite{gra2017}.

\begin{figure}[h]
\includegraphics[width=10cm]{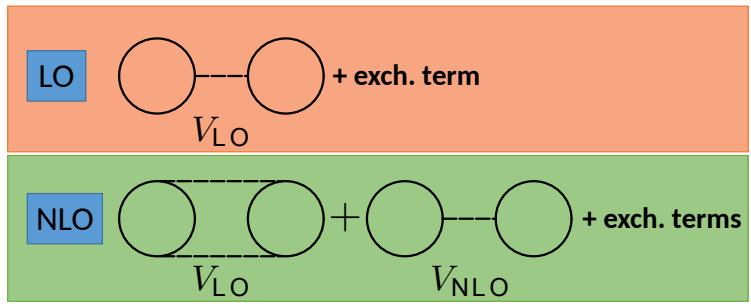}
\caption{The diagrammatic representation of contributions up to NLO for our EDF calculations.}
\label{fignlo}
\end{figure}

If one assumes that particles move in an average
mean field constructed from an effective interaction $V_{eff}$, only the upper diagram in Fig.\ref{j3} (plus the exchange term) has to 
be evaluated for the computation of the energy. 
Generally, the parameters appearing in the effective interaction $V_{eff}$ are obtained by a fit to
various nuclear properties such as binding energies and this adjustment is performed in most cases at the mean--field level. It is shown for example that a
reasonable fit can be achieved for nuclear matter and some selected nuclei 
 with a $V_{eff}$ of zero range (Skyrme--like interaction) 
or of finite range (Gogny interaction) \cite{gogny1,gogny2}. From an EFT point of view,
this indicates that:

\begin{enumerate}
\item For densities of interest ($\rho =0-0.3$ fm$^{-3}$), there might
exist an expansion to arrange diagrams in Figs. \ref{j3} and \ref{j1} order by
order individually.

\item When inserting the propagator $G$, which contains the LO contribution from Fig. \ref{j1}
into the LO diagram (Fig.1 upper part), the effect on the dressed propagator can be shifted to an
effective interaction. One can thus define for instance an LO effective interaction, $V_{\rm LO}$, which is the one used to compute the LO contribution in Fig. \ref{j1} in the dressing of the operator.  
\end{enumerate}

To further improve the functional, NLO corrections must be considered. First, such 
corrections obviously 
include the second--order contribution (NLO in the sense of the many--body perturbative expansion) computed from $V_{\rm LO}$ by using the lower diagram in Fig.\ref{j3}. In 
addition, one may define an NLO effective interaction, $V_{\rm NLO}$ (that may be associated to the 
dressing of the propagator up to NLO in Fig. \ref{j1})  
 and compute with such an interaction the energy contribution provided by the upper diagram in Fig.\ref{j3}.
This defines an expansion related to our EDF calculations whose strategy is illustrated in Fig. \ref{fignlo}. If only diagrams containing $V_{\rm LO}$ are retained in Fig. \ref{fignlo}, such an expansion for EDF will of course coincide with the many--body expansion of Fig. \ref{j3}.
By proceeding in such a way, a next--to--next--to--leading order (NNLO) correction 
may then also be obtained,  
that contains at least the third--order contribution from $V_{\rm LO}$ and the
mean--field energy contribution coming from $V_{\rm NNLO}.$ The exact form of $V_{\rm NLO}$
and $V_{\rm NNLO}$ are to be decided by renormalizability conditions and power
counting.  

In this work, where the final EOS is to be evaluated using an NLO EDF, we label the interaction as $V_{\rm LO}$ if its second--order contribution in the perturbative many--body expansion is included in the final EOS, and as $V_{\rm NLO}$ if its mean--field contribution corresponds to NLO in the functional providing the EOS. 

There are two features in our proposal. First, the parameters in
the interaction are to be renormalized at each order. Second, the $V_{eff}$
constructed in this way is specifically designed for a beyond--mean--field 
framework where the independent--particle picture on which the mean field is 
based is completely lost.
Corrections related to additional correlations such as for instance pairing correlations are 
not taken into account at the present stage. 

To establish a power counting, some assumptions are necessary here. First,
we arrange the interaction terms according to their contributions in powers of $%
k_F$ in the EOS. We denote the breakdown scale of our expansion as $\Lambda _{hi}$.   
 Then, instead of on a dilute--limit expansion 
 \cite{hammer,hammerlucas}, our power counting will be built on $\frac{k_F}{%
\Lambda _{hi}}$. We require that this expansion holds for $\rho =\rho_L- 0.4$ fm%
$^{-3}$, $\rho_L$ being the lowest density where a Skyrme--like interaction holds.  
{To guarantee that  $O\left( \left(\frac{k_F}{\Lambda _{hi}}\right)^{n+1}\right) $ contributes 
less than $O\left( \left(\frac{k_F}{\Lambda _{hi}}\right)^{n}\right) $, we fix a 
breakdown scale  $\Lambda _{hi}$ so that  $\Lambda _{hi}>k_F$. For the largest 
density that we consider for the validity of our expansion, 
$\rho=$0.4 fm$^{-3}$, $\Lambda _{hi}$ should be larger than 2.3 (1.8) fm$^{-1}$ for neutron 
(symmetric) matter. The fact that  $O\left( \left(\frac{k_F}{\Lambda _{hi}}\right)^{n+1}\right) $ 
contributes less than $O\left( \left(\frac{k_F}{\Lambda _{hi}}\right)^{n}\right) $ 
should be confirmed by analyzing the power counting. }  
 
Second, since $V_{\rm LO}$ and $V_{\rm NLO}$
are not calculated directly in this work, it is preferable to make as least
assumptions in the form of these interactions as possible. It is suggested in Ref.
  \cite{bira} that, to avoid a proliferation in the number of contact terms and at
the same time have a reasonable fit of the EOS at LO, the preferable $V_{\rm LO}$
corresponds to a $t_{0}-t_{3}$--like model. Then, throughout this work, our
strategy is to utilize RG-analysis and renormalizability-check as tools to
decide the structure of $V_{\rm NLO}$. 

\subsection{Leading order for EDF}

The simplest form of interaction at LO in the momentum space
contains $t_{0}(1+P_{\sigma }x_{0})$ only, where $t_{0}$ and $x_{0}$ are free
parameters and $P_{\sigma }=(1+\sigma _{1}\cdot \sigma _{2})/2$ is the spin--exchange 
operator. For pure neutron matter, a reasonable fit of EOS can be achieved by just one constant, that is the Bertsch parameter  \cite{bertsch}, which corresponds to the LO result from an expansion around the unitary limit \cite{denis17,denis1}\footnote{Note that the Bertsch parameter is proportional to the kinetic term rather than the $t_0$ term in the Skyrme interaction.}.
However, this interaction fails to produce
a reasonable fit for the EOS of symmetric matter at both mean--field
level and with the second--order correction included ($\chi ^{2}>1000$ for both
cases$)$ \cite{bira}. Moreover, from the study of pionless EFT, it is established that the 3-body force is required at LO to avoid the triton from collapsing \cite{3f}. This suggests that, once symmetric matter is considered, 
a three-body force 
 is required already at LO in the effective interaction. In the Skyrme case, 
 the collapse is avoided by introducing the so-called $t_3$ density--dependent two--body effective interaction. 
The next simplest form is a $t_{0}-t_{3}$--like model,
which contains a density--dependent term, that is 
\begin{eqnarray}
V_{\rm LO}&=& t_0(1+x_0P_{\sigma})
+\frac{t_3}{6} (1+x_3P_{\sigma}) \rho^{\alpha}, 
\label{vlo}
\end{eqnarray}
and gives the mean--field EOSs for symmetric and neutron matter as%
\begin{equation}
\frac{E_{SM}^{\rm (LO)}}{A}=\frac{3}{10}\frac{k_{F}^{2}}{m}+\frac{1}{4}\frac{t_{0}%
}{\pi ^{2}}k_{F}^{3}+\frac{1}{16}t_{3}\left( \frac{2}{3\pi ^{2}}\right)
^{\alpha +1}k_{F}^{3\alpha +3},  \label{meanfieldsm}
\end{equation}%
\begin{eqnarray}
\frac{E_{NM}^{\rm (LO)}}{N} &=&\frac{3}{10}\frac{k_{F}^{2}}{m}+\frac{1}{12\pi ^{2}%
}t_{0}(1-x_{0})k_{F}^{3}  \notag \\
&+&\frac{1}{24}t_{3}(1-x_{3})\left( \frac{1}{3\pi ^{2}}\right) ^{\alpha
+1}k_{F}^{3\alpha +3}.  \label{meanfieldnm}
\end{eqnarray}

Note that we adopt here natural units $\hbar=c=1$. 
The subscripts SM and NM represent symmetric and neutron matter, respectively; $t_0$, $x_0$,
$t_{3}$, $x_{3} $, and $\alpha $ are free parameters, $m$ is the nucleon mass, and    
$k_{F}=(3\pi ^{2}\rho/2)^{1/3}$ ($k_{F}=(3\pi ^{2}\rho )^{1/3}$) for SM (NM).
Note that, for $m$, one could choose to have it as an additional free
parameter in principle, as done in Ref. \cite{bira}.
Here, we adopt the point of view that all effects which modify the fermion propagator
can be transferred order by order into $V_{eff}$ as an expansion in ($%
k_{F}/\Lambda _{hi})^{n}$. Thus, the density--dependent part of the effective
mass will be encoded into our effective potential, and $m=939$ MeV is
adopted throughout this work. 

We then perform best fits to determine the free parameters ($t_0$, $t_3$, $x_0$, $x_3$, $\alpha$). The $\chi^2$ values are calculated as $%
\chi^2=\frac{1}{(N-1)}\sum_{i}\frac{(E_i-E_{i,ref})^2}{\Delta E_i^2}$, where $N$ is the
number of points on which the adjustment is done, the sum runs over this
number, $E_{i,ref}$ is the benchmark value corresponding to the point $i$,
and $\Delta E_i$ are all chosen equal to 1\% of the reference value. In this work, we take $N=10$ (ten density values from 0 to 0.3 fm$^{-3}$), we choose as benchmark EOSs the 
mean--field SLy5 EOSs \cite{sly5}, and we perform a simultaneous fit to symmetric and pure neutron matter. The $\chi ^{2}$ value is
listed in Table \ref{t1} together with the values of the parameters and the LO EOSs after fit are plotted in Fig. \ref%
{plotlo}. As we can see, both EOSs (symmetric and neutron matter) are in
quite reasonable agreement with the benchmark SLy5 mean--field curves \cite{sly5}. 
In table \ref{k}, we compare the reference SLy5 values of the saturation density $\rho_s$, the incompressibility $K_\infty$ as well 
as the saturation energy $E(\rho_s)/A$ of symmetric matter to the values obtained at LO with the minimalist $t_0$--$t_3$ model. 
Except for the incompressibility, which is slightly overestimated, the reference EOS properties 
are rather well reproduced.   
\begin{figure}[tbp]
\includegraphics[scale=0.35]{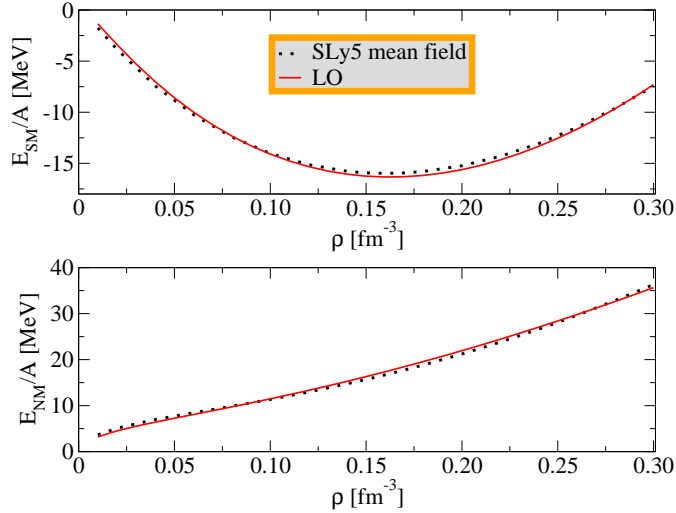}
\caption{Reference EOS as a function of the density $\protect\rho $  obtained with the SLy5 functional  (black dotted line) for symmetric (upper panel) and neutron (lower panel) matter. 
The LO EOSs (red solid line) are obtained by using Eqs. (\protect \ref{meanfieldsm}) and (\protect\ref{meanfieldnm}) with the parameters listed in Tables \protect\ref{t1}. }
\label{plotlo}
\end{figure}

\begin{table}[htbp]
\begin{center}
\begin{tabular}{cccccc}
\hline\hline
$\alpha $ & $t_{0}$ (MeV fm$^{3}$) & $t_{3}$ (MeV fm$^{3+3\alpha}$) & $x_{0}$
& $x_{3}$ & $\chi ^{2}$ \\ \hline
0.4 & -1686 & 12096 & 0.2751 & 0.2530 & 77 \\ \hline\hline
\end{tabular}
\end{center}
\caption{Parameter sets obtained by fitting the renormalized LO
EOS on the SLy5 mean--field EOS.}
\label{t1}
\end{table}

\begin{center}
\begin{table}[htbp]
\begin{center}
\begin{tabular}{|c|c|c|c|c|c|}
\hline
& SLy5 & LO & NLO$_{abc}$& NLO$_{bc}$ & NLO$_c$ \\ \hline
$\frac{E(\rho_s)}{A}$ (MeV)& -16.18 & -16.31 & -15.93 & $-15.98\pm0.1$ & $-15.97\pm0.1$ \\ \hline
$\rho_s$ (fm$^{-3}$)& 0.162 & 0.162 & 0.16 & $0.16\pm0.003$ & $0.16\pm0.003$ \\ \hline
$K_{\infty}$ (MeV)& 232.67 & 254.64 & 236.32 & $234.3\pm3.5$ & $233.2\pm3.7$ \\ \hline
\end{tabular}%
\end{center}
\caption{Saturation density $\rho_s$, saturation energy $\frac{E(\rho_s)}{A}$  and incompressibility $K_{\infty}$ for symmetric matter provided by the SLy5 mean--field EOS, the $t_0-t_3$ model (LO) and our differents scenarios for NLO with three types of counter terms (see text). }
\label{k}
\end{table}
\end{center}

\section{Next--to--leading order for EDF}

At NLO EDF one needs to consider the second--order corrections of the LO
interaction and the first--order contribution from an NLO effective interaction. 
The latter will be determined based on renormalizability and RG
analysis. 
\subsection{NLO contribution of $V_{\rm LO}$ to the EDF}
The second--order corrections (many--body perturbative expansion) in the EOS for a $t_0-t_3$ LO effective interaction
were evaluated in Refs. \cite{mog,ym,kaiser}. Here, we just report the results
relevant for our LO interaction. The second--order 
corrections consist of three parts: (a) a finite part, $\frac{%
\Delta E_{f}^{(2)}(k_F)}{A}$; (b) a divergent part with a $k_F$--dependence 
already present at LO, $\frac{\Delta E_{a}^{(2)}(k_F,\Lambda )%
}{A}$; (c) a divergent part with a $k_F$--dependence not present at LO, $%
\frac{\Delta E_{d}^{(2)}(k_F,\Lambda )}{A}$. Here, $\Lambda $ is a sharp
cutoff on the outgoing relative momentum $\vec{k}^{\prime }=(\vec{k}%
_{1}^{\prime }-\vec{k}_{2}^{\prime })/2$, with $\vec{k}_{1,2}^{(\prime )}$
being the incoming (outgoing) momentum of nucleon $1,2$. For symmetric
matter, the second--order correction reads 
\begin{equation}
\frac{\Delta E_{SM,f}^{(2)}(k_{F})}{A}=\frac{3m}{\pi ^{4}}\frac{\left[11-2\ln 2 \right]}{%
280}k_{F}^{4}\left[ 
\begin{array}{c}
t_{0}^{2}\left( 1+x_{0}^{2}\right) +2t_{0}T_{3}\left( 1+x_{0}x_{3}\right)
k_{F}^{3\alpha }+(1+x_{3}^{2})T_{3}^{2}k_{F}^{6\alpha }+ \\ 
\frac{3}{8}t_{0}T_{3}k_{F}^{3\alpha }\alpha (3+\alpha )+\frac{3}{8}%
T_{3}^{2}k_{F}^{6\alpha }\alpha (3+\alpha )+\frac{9}{256}T_{3}^{2}k_{F}^{6%
\alpha }\alpha ^{2}(3+\alpha )^{2}%
\end{array}%
\right] ,  \label{terms1}
\end{equation}

\begin{equation}
\frac{\Delta E_{SM,a}^{(2)}(k_{F},\Lambda )}{A}=-\frac{m}{8\pi ^{4}}\Lambda
k_{F}^{3}\left[ t_{0}^{2}\left( 1+x_{0}^{2}\right) +2t_{0}T_{3}\left(
1+x_{0}x_{3}\right) k_{F}^{3\alpha }+\frac{3}{8}t_{0}T_{3}\alpha (3+\alpha
)k_{F}^{3\alpha }\right] ,  \label{terms2}
\end{equation}

\begin{equation}
\frac{\Delta E_{SM,d}^{(2)}(k_{F},\Lambda )}{A}=-\frac{m}{8\pi ^{4}}\Lambda
k_{F}^{3+6\alpha }T_{3}^{2}\left[ (1+x_{3}^{2})+\frac{9}{256}\alpha
^{2}(\alpha +3)^{2}+\frac{3}{8}\alpha (\alpha +3)\right] ,  \label{terms3}
\end{equation}%
where%
\begin{equation}
T_{3}=\left(\frac{2}{3\pi ^{2}}\right)^{\alpha }\frac{t_{3}}{6}.
\end{equation}

For neutron matter, one has%
\begin{equation}
\frac{\Delta E_{NM,f}^{(2)}(k_{F})}{A}=\frac{m}{\pi ^{4}}\frac{\left[11-2\ln 2\right]}{280%
}k_{F}^{4}\left[ (T_{0}+k_{F}^{3\alpha }T_{3}^{R})^{2}\right] ,  \label{n1}
\end{equation}

\begin{equation}
\frac{\Delta E_{NM,a}^{(2)}(k_{F},\Lambda )}{A}=-\frac{m}{24\pi ^{4}}\Lambda
k_{F}^{3}\left[ T_{0}^{2}+2T_{0}T_{3}^{R}k_{F}^{3\alpha }\right] ,
\label{n2}
\end{equation}

\begin{equation}
\frac{\Delta E_{NM,d}^{(2)}(k_{F},\Lambda )}{A}=-\frac{m}{24\pi ^{4}}\Lambda
k_{F}^{3+6\alpha }\left[ (T_{3}^{R})^{2}\right] ,  \label{n3}
\end{equation}

\bigskip where%
\begin{eqnarray}
T_{0} &=&t_{0}(1-x_{0}) \\
T_{3}^{R} &=&\left(\frac{1}{3\pi ^{2}}\right)^{\alpha }\left[ \frac{t_{3}}{6}(1-x_{3})+%
\frac{1}{48}t_{3}(1-x_{3})\alpha (3+\alpha )\right] .
\end{eqnarray}%
The contribution from the rearrangement terms \cite{car,pastore} is included in the above
equations. A summary of the different $k_F$ dependences in the EOS is shown in table \ref{kflambda}.

\begin{table}[htbp]
\centering%
\begin{tabular}{cllll}
\hline\hline
Contribution to $E/A$ & \hspace*{1.cm} $V_{\rm LO}$  \hspace*{1.cm} & \hspace*{1.cm} $V^{(a)}_{\rm NLO}$ \hspace*{1.cm}&
\hspace*{1.cm}  $V^{(b)}_{\rm NLO}$  \hspace*{1.cm} & \hspace*{1.cm} $V^{(c)}_{\rm NLO}$ \hspace*{1.cm}
 \\ \hline
 Mean field &  \hspace*{1.cm} $ k^{3}_F$, ~$k^{3+3\alpha}_F$ &  \hspace*{1.cm} \textcolor{blue}{$ k^{3}_F$} 
 & \hspace*{1.cm} \textcolor{blue}{$ k^{3+3\alpha}_F$}  
 & \hspace*{1.cm}  \textcolor{red}{$k^{3+6\alpha}_F$}   \\ \hline
 Second--order &  \hspace*{1.cm} $k^{4}_F$, ~$k^{4+3\alpha}_F$,~$k^{4+6\alpha}_F$ \\
 & \hspace*{1.cm} \textcolor{blue}{$ k^{3}_F$,  $ k^{3+3\alpha}_F$} \\
 &  \hspace*{1.cm} \textcolor{red}{$k^{3+6\alpha}_F$}
  \\ \hline\hline
\end{tabular}%
\caption{Different $k_F$ 
 dependences in the EOSs of neutron and symmetric matter obtained for different interactions discussed in the text. All the terms in black are cutoff independent. In the second--order contribution of $V_{\rm LO}$, the terms in red and blue are linearly cutoff dependent. 
In particular, the terms in blue 
can be treated either by absorbing them in the mean--field part by a redefinition of the 
parameters or 
by introducing counter terms of the type $V_{\rm NLO}^{(a)}$ and $V_{\rm NLO}^{(b)}$. The terms in red in the second--order contribution correspond, in the absence of restrictions on the $\alpha$ values, to terms that require explicitly the introduction of counter terms, $V_{\rm NLO}^{(c)}$. 
 Note finally that the second--order contributions of $V_{\rm NLO}$
are not shown since they will appear only when going to higher orders. }
\label{kflambda}
\end{table} 

\begin{figure}[htbp]
\includegraphics[scale=0.35]{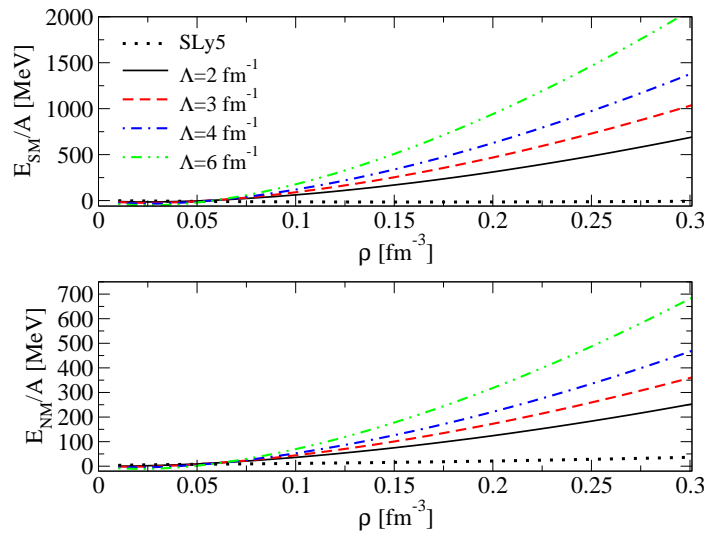}
\caption{EOS as a function of the density $\protect\rho $ 
for symmetric (upper panel) and neutron (lower panel) matter. The
dotted line represents the mean--field SLy5 EOSs. The LO parameters listed in Table \ref{t1} 
are used to compute the EOSs generated from the NLO EDF using $V_{\rm LO}$. }
\label{unren}
\end{figure}

\subsection{Scenarios for regularization}

In Fig. \ref{unren} we plot the unrenormalized EOSs obtained by including the contributions generated from the NLO EDF using the $V_{\rm LO}$ interaction, where we simply use the LO parameters listed in Table \ref{t1}. As one can see, both EOSs show strong cutoff dependence and, as the increase of $\Lambda$, depart further away from the benchmark value. This shows that renormalization is required.    

The separation of the second--order part into three contributions is at the heart of the strategy 
we use below to propose different scenarios for regularisation. Let us first start with preliminary remarks that are important for 
the coming discussion:
\begin{itemize}
  \item Except for the finite parts (Eqs. (\ref{terms1}) and (\ref{n1})), the exact forms of $\frac{\Delta E_{a}^{(2)}}{A}$ and $%
\frac{\Delta E_{d}^{(2)}}{A}$ (Eqs. (\ref{terms2})-(\ref{terms3}) and Eqs. (\ref{n2})-(\ref{n3})) are regularization--scheme dependent. However, for our expansion to make sense, the final EOS should not depend on a particular scheme after proper renormalization. This will be verified in the following by comparing the effect of various counter terms.   
  \item The parameter $\alpha $, which appears in the density--dependent term, requires
a special attention because each value of $\alpha$ would provide a different $k_F$ dependence. 
In the present work, we keep $\alpha $ as a free parameter in
the renormalization.
  \item The highest $k_F$--dependence appearing in the second--order
EOS is $k_F^{4+6\alpha }$. Thus, by a simple counting in powers of $k_F$,
the $t_{1}$ and $t_{2}$ terms of the Skyrme interaction (which contribute at first order as $k_F^{5}$ in the 
EOS) do not enter in the NLO effective interaction for $\alpha <\frac{1}{6}$.  In the following, since $\alpha$ is varied freely, $\alpha$ might 
exceed $\frac{1}{6}$. In this case, one should keep in mind that, a priori, one has also to include the $t_1$ and $t_2$ terms of the Skyrme 
interaction.
\end{itemize}
In a previous study \cite{bira}, it was shown that the divergence appearing in  $\frac{\Delta E_{a}^{(2)}}%
{A}$ may be absorbed by a redefinition of the existing parameters since those terms have the same 
$k_F$ dependence as in first--order terms. For the divergence appearing in 
$\frac{\Delta E_{d}^{(2)}}{A}$, one could first
search for some special values of $\alpha $ which would give for $\frac{\Delta
E_{d}^{(2)}}{A}$ the same $k_F-$dependences as those
appearing in the mean--field part. Then, one could perform the renormalization by absorbing
the $\Lambda -$divergence into a redefinition of the parameters. This approach
was adopted in Ref. \cite{bira}, where no new counter terms were included. 

 In this work, we adopt a more general approach. We release the
requirement on specific values of $\alpha $ and, in general, we allow treating 
$\frac{\Delta E_{a}^{(2)}%
}{A}$  and $\frac{\Delta E_{d}^{(2)}}{A}$ in the same way: 
both divergences present in 
$\frac{\Delta E_{a}^{(2)}}{A}$ and $\frac{\Delta
E_{d}^{(2)}}{A}$ may be directly renormalized by NLO EDF
contributions. This allows us to use the divergence generated at NLO by 
an LO interaction  
as an important guide
for the construction of an NLO effective interaction, denoted by $V_{\rm NLO}$. In principle, each $%
\Lambda k_F^{n}-$divergence in the EOS can be directly associated with an NLO
counter part $A_{n}k_F^{n}$, where $A_{n}$ denotes an additional free
parameter\footnote{A recent approach which constructs the interaction directly on a particular power series of $\sum_n k_F^n$ is introduced in Refs. \cite{kids,kids2}. However, in present work we consider $n$ to be any real number.}. 
A term in 
the effective interaction of the 
form of $O\left( (\vec{k}-\vec{k}^{\prime
})^{n-3v-3}\rho ^{v}\right) $ will contribute as $O\left( k_F^{n}\right) $
in the EOS, where $v$ is an arbitrary number which satisfies $n-3v-3=$even number. Note that the parameter $v$ does not appear in the EOS of matter. This additional free parameter may eventually  be adjusted with a fit   
 to reproduce properties of finite nuclei. 
 Interactions of the above type
appear naturally for example when one expands the terms coming from a resumed expression 
 \cite{steele,yglo,schafer,kaiser1}. 

Without fixing $\alpha$ to specific values, the minimum counter term required to absorb the divergences
present in $\frac{\Delta E_{d}^{(2)}}{A}$ is the one
proportional to $k_F^{3+6\alpha }$. On the other hand, the divergence
present in $\frac{\Delta E_{a}^{(2)}}{A}$ can be absorbed
by just a redefinition of $t_{0}$, $x_{0}$, $t_{3}$, and $x_{3}$ or by adding more
counter terms proportional to $k_F^{3}$ and $k_F^{3+3\alpha }$. Note that, in
both cases, the mean--field values of the parameters are modified.

The three contact interactions which correspond to the three divergences appearing $\frac{\Delta E_{a}^{(2)}}{A}$ and $\frac{\Delta E_{d}^{(2)}}{A}$ can be written as  
\begin{eqnarray}
V_{\rm NLO}^{(a)}&=&a(1+P_{\sigma }x_{a})f_a\left((\vec{k}-\vec{k}^{\prime
})^{-3v_a},\rho ^{v_a}\right),  \label{ca} \\
V_{\rm NLO}^{(b)}&=&b(1+P_{\sigma }x_{b})f_b\left((\vec{k}-\vec{k}^{\prime
})^{3\alpha-3v_b},\rho ^{v_b}\right),  \label{cb} \\
V_{\rm NLO}^{(c)}&=&c(1+P_{\sigma }x_{c})f_b\left((\vec{k}-\vec{k}^{\prime
})^{6\alpha-3v_c},\rho ^{v_c}\right),  \label{cc}
\end{eqnarray}
where $f_{a,b,c}$ are functions which contain infrared regulators to prevent potential singularities at $\rho\rightarrow 0$ or $|\vec{k}-\vec{k'}|\rightarrow 0$; it may turn out that a best fit to finite nuclei would provide negative powers for $(\vec{k}-\vec{k}^{\prime
})$ or $\rho$. Away from singularities, we have $f_{a,b,c}\left((\vec{k}-\vec{k}^{\prime
})^{n-3v-3},\rho ^{v}\right)\approx (\vec{k}-\vec{k}^{\prime
})^{n-3v-3}\rho ^{v}$. $a,$ $b$, $c$, $x_{a},$ $x_{b},$ $x_{c}$ are free parameters to be determined by an adjustment of the EOS. 
On the other hand, $v_a, v_b$ and $v_c$ are extra parameters that could be determined only through further adjustments done for finite nuclei.
With their mean--field contribution directly entering in the NLO EOS, the above three counter terms provide $k_F^3$, $k_F^{3\alpha+3}$, and $k_F^{6\alpha+3}$ terms to the EOS (see table \ref{kflambda}). 
Note that only Eq. (\ref{cc}) (with contribution $k_F^{6\alpha+3}$) is necessarily required by renormalizability. The effect of the other two terms (Eqs. (\ref{ca}) and (\ref{cb})) can be replaced by readjusting the values of $t_{i's}$ and $x_{i's}$ so that these two counter terms, for nuclear matter, should just modify the values of the parameters and not the power counting. 

In table \ref{list}, we list all Skyrme--type $V_{\rm LO}$ and $V_{ \rm NLO}$ interactions and the $k_F$ dependencies generated in the EOS from the LO and NLO EDFs. We show in red the $V_{ \rm NLO}$ contributions which are not included in the present study because we limit $\alpha$ to be less than 1/6.

\begin{table}[htbp]
\begin{center}
\begin{tabular}{|c|c|c|}
\hline\hline
Skyrme--type interaction & $k_F$-dep. in the EOS from LO EDF & $k_F$-dep. in the EOS from NLO EDF  \\ \hline
$V_{\rm LO}$: $t_0(1+x_0P_{\sigma})$ & $k_F^3$ & $t_0^2$ terms: $k_F^3$, $k_F^4$ \\ \hline
$V_{\rm LO}$: $t_3(1+x_3P_{\sigma})\rho^{\alpha}$ & $k_F^{3+3\alpha}$ & $t_3^2$ terms: $k_F^{3+6\alpha}$, $k_F^{4+6\alpha}$ \\ \hline
                                         &                   & $t_0t_3$ terms:  $k_F^{3+3\alpha}$, $k_F^{4+3\alpha}$ \\ \hline
$V_{\rm NLO}$ (counter terms): Eq. (13)                           &                   &  $k_F^3$ \\ \hline
$V_{\rm NLO}$ (counter terms): Eq. (14)                           &                   &  $k_F^{3+3\alpha}$ \\ \hline
$V_{\rm NLO}$ (counter terms): Eq. (15)                           &                   & $k_F^{3+6\alpha}$ \\ \hline
\textcolor{red}{$V_{\rm NLO}$: $t_1(1+x_1P_{\sigma})(\mathbf{k%
}^{\prime 2}+\mathbf{k}^{2})$} &  & \textcolor{red}{$k_F^5$} \\ \hline
\textcolor{red}{$V_{\rm NLO}$: $t_2(1+x_2P_{\sigma})\mathbf{k}^{\prime
}\cdot \mathbf{k}$} &    & \textcolor{red}{$k_F^5$} \\ \hline
\end{tabular}
\end{center}
\caption{Skyrme--type $V_{\rm LO}$ and $V_{\rm NLO}$ interactions and $k_F$ dependencies generated in the EOS from the LO and NLO EDFs. We show in red the $V_{\rm NLO}$ contributions which are not included in the present study because we limit $\alpha$ to be less than 1/6. Note that, here, we do not include spin-orbit and tensor interactions  
that should a priori appear as $V_{\rm NLO}$ and contribute at EDF NLO with their mean-field functional. The reason is that such mean-field contributions are zero in infinite matter.}
\label{list}
\end{table}

The scenario we consider for regularization 
will depend on the type of counter terms that are included in $V_{\rm NLO}$. Since the case with no counter term has already been 
discussed in Ref. \cite{bira}, we consider three possible scenarios, referred as scenario (c), (bc) and (abc) that refers to the fact that 
only $V^{(c)}_{\rm NLO}$, only  $V^{(b)}_{\rm NLO}$ plus $V^{(c)}_{\rm NLO}$, or all three counter terms are used to construct $V_{\rm NLO}$, respectively. The resulting  
EOSs will be respectively called EOS-NLO${_c}$, EOS-NLO${_{bc}}$ and EOS-NLO${_{abc}}$.

 {\bf Scenario (abc):} Adopting all three types of counter terms, the EOS up to
NLO reads%
\begin{eqnarray}
\frac{E_{SM}^{\rm (NLO)}(k_{F})}{A} &=&\frac{3}{10}\frac{k_{F}^{2}}{m}+\frac{%
k_{F}^{3}}{4\pi ^{2}}[t_{0}+A]+\frac{k_{F}^{3\alpha +3}}{4\pi ^{2}}[T_{3}+B]-%
\frac{m}{8\pi ^{4}}k_{F}^{3+6\alpha }C + 
\frac{\Delta E_{SM,f}^{(2)}(k_{F})}{A} 
  \label{abc}
\end{eqnarray}
for symmetric matter and 
\begin{eqnarray}
\frac{E_{NM}^{\rm (NLO)}(k_{F})}{A} &=&\frac{3}{10}\frac{k_{F}^{2}}{m}+%
\frac{1}{12\pi ^{2}}\left[ t_{0}(1-x_{0})+A^{\ast }\right] k_{F}^{3}+\left[ 
\frac{1}{24}t_{3}(1-x_{3})\left( \frac{1}{3\pi ^{2}}\right) ^{\alpha +1}+%
\frac{B^{\ast }}{4\pi ^{2}}\right] k_{F}^{3\alpha +3}  \notag \\
&&-\frac{m}{8\pi ^{4}}k_{F}^{3+6\alpha }C^{\ast } + 
\frac{\Delta E_{NM,f}^{(2)}(k_{F})}{A}
\label{abc2}
\end{eqnarray}%
for neutron matter. 
Note that, to simplify the notation, we have defined $A^{(\ast )}$, $%
B^{(\ast )}$, and $C^{(\ast )}$ as the parameters originating from Eqs.~(\ref%
{ca}), (\ref{cb}), and (\ref{cc}) for symmetric
(neutron) matter. The parameters $a$, $b$ and $c$ in Eqs.~(\ref%
{ca}), (\ref{cb}), and (\ref{cc}) can be splitted into two parts. One cancels the linear ($\Lambda$) divergence in the EOS. The remaining parts are finite, are denoted by $A^{(\ast )}$, $%
B^{(\ast )}$, and $C^{(\ast )}$ and enter into the fitting procedure. 

\begin{table}[tbp]
\centering%
\begin{tabular}{cccccccccccc}
\hline\hline
$\alpha $ & $t_{0}$ (MeV fm$^{3}$) & $t_{3}$ (MeV fm$^{3+3\alpha}$) & $x_{0}$
& $x_{3}$ & $A$(MeV fm$^{3}$) & $B$ (MeV fm$^{3+3\alpha}$) & $C$ (MeV fm$%
^{3+6\alpha}$) &  &  &  &  \\ \hline
-0.083 & 307.6 & 97.27 & -2.721 & -13.31 & -7329 & 8339 & 14965 &  &  &  &
\\ \hline\hline
$A^{\ast}$(MeV fm$^{3}$) & $B^{\ast }$(MeV fm$^{3+3\alpha}$) & $C^{\ast }$
(MeV fm$^{3+6\alpha}$) & $\chi ^{2}$ &  &  &  &  &  &  &  &  \\ \hline
-24149 & 11159 & 18781 & 0.46 &  &  &  &  &  &  &  &  \\ \hline\hline
\end{tabular}%
\caption{Parameter sets obtained by fitting the renormalized second--order
EOS to the SLy5 mean--field EOS. Here the second--order EOSs reported in Eqs. (%
\protect\ref{abc}) and (\protect\ref{abc2}) are used.}
\label{tnloa}
\end{table} 

\begin{figure}[htbp]
\includegraphics[scale=0.35]{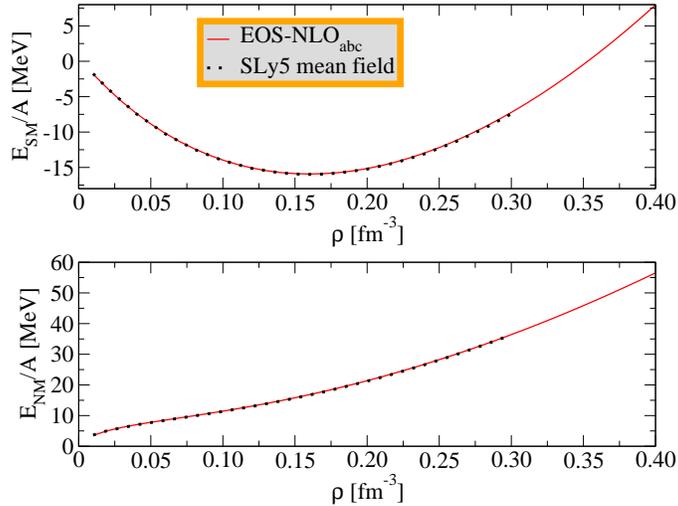}
\caption{EOS as a function of the density $\protect\rho $,
where the subscript SN (NM) represents symmetric (neutron) matter. The
dotted (red solid) line represents the mean--field SLy5 (renormalized
NLO) EOSs. The NLO EOSs are obtained
by using the scenario (abc), that is  Eqs. (\protect\ref{abc}) and (\protect\ref{abc2}), with the parameters listed
in Table \protect\ref{tnloa}. }
\label{plotnloa}
\end{figure}

No cutoff is present in Eqs.~(\ref{abc}) and (\ref{abc2}) because 
all possible divergences have been absorbed by counter
terms. We then perform the renormalization by refitting Eqs.~(\ref{abc}) and (\ref{abc2}) to a
 benchmark symmetric and neutron matter EOS, given by the SLy5 Skyrme interaction at the 
mean--field level, from $\rho=0\sim0.3$ fm$^{-3}$. In Fig. \ref{plotnloa}, we plot the resulting EOS for symmetric and
neutron matter up to $\rho=0.4$ fm$^{-3}$. As one can see, both the fit in
symmetric and neutron matter agree with the standard value with $\chi ^{2}=0.46$
as listed in Table \ref{tnloa}. However, it is not possible to perform
a RG-analysis in this case because no cutoff-dependence is present in the final
EOS.

 {\bf Scenario (bc):} 
Next, we renormalize the second--order EOS in the absence of the $a-$counter term (Eq. (\ref{ca})) and let the $k_F^3-$divergence be absorbed by a redefinition of the parameters.
The resulting EOS reads%
\begin{eqnarray}
\frac{E_{SM}^{\rm (NLO)}(k_{F},\Lambda)}{A} &=&\frac{3}{10}\frac{k_{F}^{2}}{m}+\frac{%
k_{F}^{3}}{4\pi ^{2}}t_{0}^{\Lambda}+\frac{k_{F}^{3\alpha +3}}{4\pi ^{2}}[T_{3}+B]-%
\frac{m}{8\pi ^{4}}k_{F}^{3+6\alpha }C  + \frac{\Delta E_{SM,f}^{(2)}(k_{F})}{A}
  \label{bc}
\end{eqnarray}
for symmetric matter and 
\begin{eqnarray}
\frac{E_{NM}^{\rm (NLO)}(k_{F},\Lambda)}{A} &=&\frac{3}{10}\frac{k_{F}^{2}}{m}+%
\frac{k_{F}^{3}}{12\pi ^{2}}t_{0}^{\Lambda}(1-x_{0}^{\Lambda})+\left[ \frac{1}{24}%
t_{3}(1-x_{3})\left( \frac{1}{3\pi ^{2}}\right) ^{\alpha +1}+\frac{B^{\ast }%
}{4\pi ^{2}}\right] k_{F}^{3\alpha +3}  \notag \\
&&-\frac{m}{8\pi ^{4}}k_{F}^{3+6\alpha }C^{\ast }+\frac{\Delta E_{NM,f}^{(2)}(k_{F})}{A}
\label{bc2}
\end{eqnarray}
for neutron matter. 
Here, $t_0^{\Lambda}$ and $x_0^{\Lambda}$ are
\begin{eqnarray}
t_{0}^{\Lambda} &=&t_{0}-\frac{m\Lambda }{2\pi ^{2}}t_{0}^{2}\left(
1+x_{0}^{2}\right) ,  \label{t0r} \\
t_{0}^{\Lambda}(1-x_{0}^{\Lambda}) &=&t_{0}(1-x_{0})-\frac{m\Lambda }{2\pi ^{2}}%
t_{0}^{2}\left( 1-x_{0}\right) ^{2} . \label{t3r}
\end{eqnarray}
Note that, through $t_0^{\Lambda}$ and $x_0^{\Lambda}$, $\Lambda$ is present in Eqs. (\ref{bc}) and (\ref{bc2}). However, together with Eqs. (\ref{t0r}) and (\ref{t3r}), it is clear that the cutoff dependence in the final EOS can always be eliminated properly\footnote{After renormalization, one is left with a residual cutoff-dependence of the order $\Re(\Lambda,k_F,\Lambda_{hi})\left(\frac{k_F}{\Lambda_{hi}}\right)^{n+1}$, where $\Re$ is a function of the natural size and $n$ is the order of the calculation \cite{gries}.} after the renormalization is done. We then perform again a best
fit to the mean--field SLy5 EOS (from $\rho=0\sim0.3$ fm$^{-3}$), for $\Lambda =1.2-20$ fm$^{-1}$. The
resulting EOSs for symmetric and neutron matter are plotted in Fig.\ref%
{plotnlob}, and the parameters and corresponding $\chi ^{2}$ are listed
in Table \ref{tnlob}.

\begin{table}[htbp]
\begin{tabular}{|c|c|c|c|c|c|c|c|c|c|c|}
\hline
$\Lambda $ (fm$^{-1}$) & 2 & 4 & 6 & 8 & 10 & 12 & 14 & 16 & 18 & 20 \\ \hline
$t_{0}$ (fm$^{2}$) & -2.804 & 2.024 & -1.146 & 4.443 & 1.415 & 1.206 & -1.960 & 2.627 & -0.6473 & 0.4415\\ \hline
$t_{3}$ (fm$^{2+3\alpha }$) & 31.89 & -28.99 & -4.159 & -47.48 & 2.661 & 18.31 & -15.53 & -0.5724 & -20.38 & -21.44
\\ \hline
$x_{0}$ & -2.229 & 1.350 & 1.095 & 0.6359 & 1.448 & 1.202 & 1.203 & -0.1834 & 4.257 & 0.3196\\ \hline
$x_{3}$ & -1.463 & 2.059$\cdot 10^{-3}$ & -6.376 & 0.1812 & -11.70 & -0.7088 & -0.9565 & 31.64 & -1.103 & -0.4495\\ 
\hline
$B$ (fm$^{2+3\alpha }$) & 14.54 & -29.11 & -23.61 & 65.71 & -4.749 & -16.44 & 50.99 & 29.71 & 39.75 & -28.14\\ 
\hline
$C$ (fm$^{2+6\alpha }$) & -2.713 & -146.3 & -93.89 & 67.63 & -46.00 & -59.68 & 97.23 & 51.99 & 37.24 & -104.2\\ 
\hline
$B^{\ast }$ (fm$^{2+3\alpha }$) & 28.67 & 37.49 & 10.77 & 17.46 & 7.152 & 19.17 & 4.072 & 25.15 & 32.86 & 8.702
\\ \hline
$C^{\ast }$ (fm$^{2+6\alpha }$) & 73.58 & 160.3 & 49.08 & 69.76 & 44.66 & 96.88 & 20.12 & 73.83 & 114.1 & 39.76
\\ \hline
$\alpha $ & 4.77$\cdot 10^{-2}$ & 1.48$\cdot 10^{-2}$ & 3.13$\cdot 10^{-2}$
& 2.28$\cdot 10^{-2}$ & 3.59$\cdot 10^{-2}$ & 1.68$\cdot 10^{-2}$ & 4.96$\cdot 10^{-2}$ & 6.48$\cdot 10^{-2}$ & 1.92$\cdot 10^{-2}$ & 3.44$\cdot 10^{-2}$  \\ \hline
$\chi ^{2}$ & 0.39 & 2.19 & 0.76 & 0.88 & 2.41 & 4.04 & 1.95 & 3.62 & 1.67 & 1.18 \\ \hline
\end{tabular}%
\caption{Parameter sets obtained by fitting the renormalized second--order
EOS to the SLy5 mean--field EOS. Here the second--order EOSs reported in Eqs. (%
\protect\ref{bc}) and (\protect\ref{bc2}) are used.}
\label{tnlob}
\end{table}

\begin{figure}[tbp]
\includegraphics[scale=0.35]{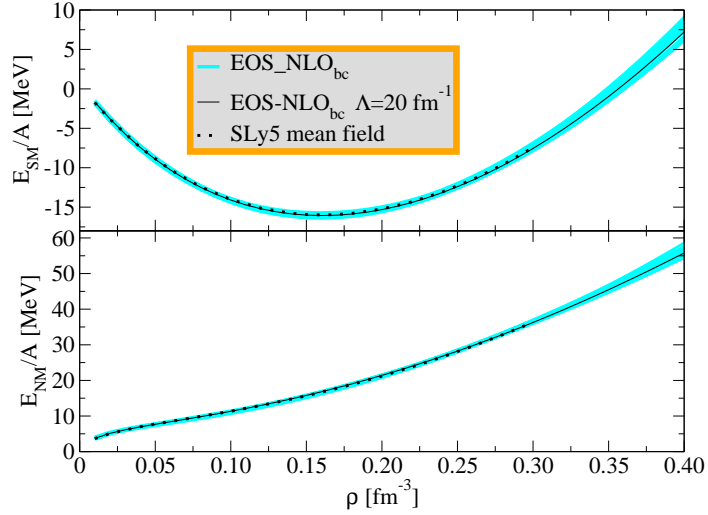}
\caption{(EOS as a function of the density $\protect\rho $,
where the subscript SN (NM) represents symmetric (neutron) matter. The
dotted line (blue band) represents the mean--field SLy5 (renormalized
NLO) EOSs. The NLO EOSs are obtained
by using the scenario (bc), that is Eqs. (\protect\ref{bc}) and (\protect\ref{bc2}) with the parameters listed in
Table \protect\ref{tnlob}. Here, the cutoff is taken in the window $\Lambda =1.2-20$ fm$^{-1}$ and the errorbars correspond to the 
cutoff dependence of the fit.}
\label{plotnlob}
\end{figure}

\begin{figure}[tbp]
\includegraphics[scale=0.35]{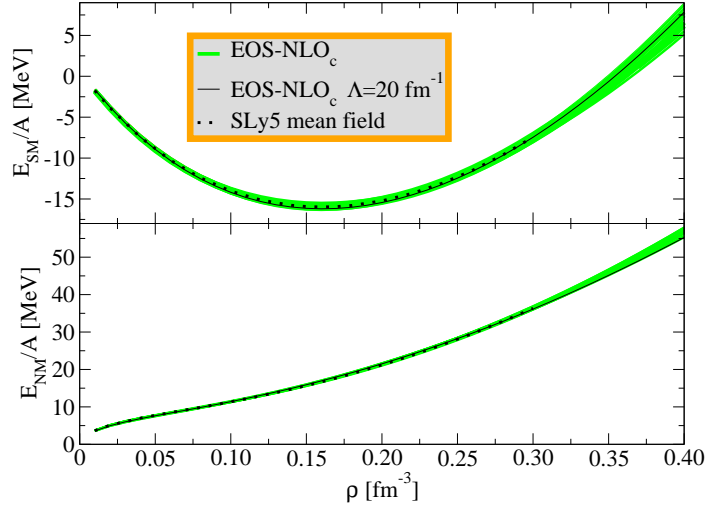}
\caption{EOSs as a function of the density $\protect\rho $,
where the subscript SN (NM) represents symmetric (neutron) matter. The
dotted line (green band) represents the mean--field SLy5 (renormalized
NLO) EOSs. The NLO EOSs are obtained
by using the scenario (c), that is Eqs. (\protect\ref{a}) and (\protect\ref{a2}) with the parameters listed in
Table \protect\ref{tnloc}. Here, the cutoff is taken in the window $\Lambda =1.2-20$ fm$^{-1}$ and the errorbars correspond to the 
cutoff dependence of the fit.}
\label{plotnloc}
\end{figure}

 {\bf Scenario (c):} 
For the case where one only allows the minimum counter term to enter, that is the counter term in Eq.~(%
\ref{cc}), the divergences in powers of $k_F^3$ and $k_F^{3\alpha+3}$ are
absorbed into a redefinition of $t_0$, $x_0$, $t_3$, and $x_3$, and the resulting
EOS reads%
\begin{eqnarray}
\frac{E_{SM}^{\rm (NLO)}(k_{F},\Lambda)}{A} &=&\frac{3}{10}\frac{k_{F}^{2}}{m}+\frac{%
k_{F}^{3}}{4\pi ^{2}}t_{0}^{\Lambda}+\frac{k_{F}^{3\alpha +3}}{4\pi ^{2}}T_{3}-%
\frac{m}{8\pi ^{4}}k_{F}^{3+6\alpha }C  \notag \\
&&-\frac{m}{8\pi ^{4}}\Lambda k_{F}^{3+3\alpha }t_{0}\left[ 2T_{3}\left(
1+x_{0}x_{3}\right) +\frac{3}{8}T_{3}\alpha (3+\alpha )\right]  + 
\frac{\Delta E_{SM,f}^{(2)}(k_{F})}{A}
  \label{a}
\end{eqnarray}
for symmetric matter and 
\begin{eqnarray}
\frac{E_{NM}^{\rm (NLO)}(k_{F},\Lambda)}{A} &=&\frac{3}{10}\frac{k_{F}^{2}}{m}+%
\frac{k_{F}^{3}}{12\pi ^{2}}t_{0}^{\Lambda}(1-x_{0}^{\Lambda})+\frac{k_{F}^{3\alpha +3}}{%
24}t_{3}(1-x_{3})\left( \frac{1}{3\pi ^{2}}\right) ^{\alpha +1}  \notag \\
&&-\frac{m\Lambda k_{F}^{3+3\alpha }}{12\pi ^{4}}T_{0}T_{3}^{R}-\frac{m}{%
8\pi ^{4}}k_{F}^{3+6\alpha }C^{\ast }+
\frac{\Delta E_{NM,f}^{(2)}(k_{F})}{A}
  \label{a2}
\end{eqnarray}%
for neutron matter. 
Here, only the $C^{(\ast )}$ counter term enters into play. Again, renormalizability is guaranteed as the divergences can always be absorbed into a redefinition of the Skyrme parameters. With the same renormalization strategy as for the previous two cases, the resulting EOSs for
symmetric and neutron matter are plotted in Fig. \ref{plotnloc}, and the
parameters and corresponding $\chi ^{2}$ are listed in Table \ref{tnloc}%
. Note that for $\Lambda > 4$ fm$^{-1}$, some of the values
of $\alpha$ exceed $1/6$. In principle, one should thus include the mean--field
contributions from the $t_{1},$ $t_{2}$--terms in the EOS in these cases. We
have performed such calculations and found that including non-zero $t_{1},$ $%
t_{2}-$terms does not improve the overall quality of the fits.
\begin{table}[tbp]
\begin{tabular}{|c|c|c|c|c|c|c|c|c|c|c|}
\hline
$\Lambda $ (fm$^{-1}$) & 2 & 4 & 6 & 8 & 10 & 12 & 14 & 16 & 18 & 20 \\ \hline
$t_{0}$ (fm$^{2}$) & -2.987 & -2.543 & -2.140 & -2.543 & -1.725 & 1.885 & -0.7193 & -1.362 & 1.512 & 1.407\\ 
\hline
$t_{3}$ (fm$^{2+3\alpha }$) & 19.36 & -0.9911 & -4.586 & 
20.02 & 0.8675 & 1.645 & 10.39 & -0.4202 & 0.7416 & 2.174\\ \hline
$x_{0}$ & 1.291 & 0.6370 & 0.8911 & 0.7149 & 0.5508 & 0.5239 & 2.247 & 0.6074 & 0.4943 & 0.5695 \\ \hline
$x_{3}$ & -0.1825 & -15.17 & -4.145 & -0.6774 & 12.87 & -7.037 & -6.189$\cdot 10^{-3}$ & -25.80 & -12.45 & -4.917\\ 
\hline
$C$ (fm$^{2+6\alpha }$) & 12.01 & -3.791 & 1.962 & 20.95 & -1.383 & -3.364 & 0.3387 & -0.8847 & -3.547 & -3.286
\\ \hline
$C^{\ast }$ (fm$^{2+6\alpha }$) & 10.64 & 61.689 & 0.7346 & 33.21 & -2.972 & 
-3.140 & -3.027 & -3.430 & -3.773 & -3.953\\ \hline
$\alpha $ & 6.88$\cdot 10^{-2}$ & 0.170 & 0.126 & 4.20$\cdot 10^{-2}$ & 
0.224 & 0.187 & 0.226 & 0.223 & 0.205 & 0.190\\ \hline
$\chi ^{2}$ & 3.01 & 1.06 & 1.25 & 2.53 & 0.34 & 0.55 & 1.63 & 0.55 & 1.92 & 1.23 \\ \hline
\end{tabular}%
\caption{Parameter sets obtained by fitting the renormalized second--order
EOS to the SLy5 mean--field EOS. Here the second--order EOSs reported in Eqs. (%
\protect\ref{a}) and (\protect\ref{a2}) are used.}
\label{tnloc}
\end{table}

So far, we have checked three out of the four possible scenarios for the NLO contact terms. We do not consider the possibility of having an $(ac)$ scenario because the NLO EOS is unlikely to consist of counter terms proportional to $k_F^3$ and $k_F^{3+6\alpha}$, without the intermediate term $k_F^{3+3\alpha}$.  

From the fact that satisfactory fits (with similar quality) can be obtained by all three scenarios, we conclude that the regularization-scheme dependence present in Eqs. (\ref{terms2})-(\ref{terms3}) and Eqs. (\ref{n2})-(\ref{n3}) does not affect the NLO results after renormalization. The differences due to the regularization scheme can be transferred into the counter terms present in Eqs. (\ref{ca}) and (\ref{cb}). 
The independence of the final result of the regularization scenario is also illustrated in Table \ref{k} where 
we see that the properties of symmetric matter are almost independent of the scenario and well match the reference SLy5 EOS.

\begin{figure}[tbp]
\includegraphics[scale=0.35]{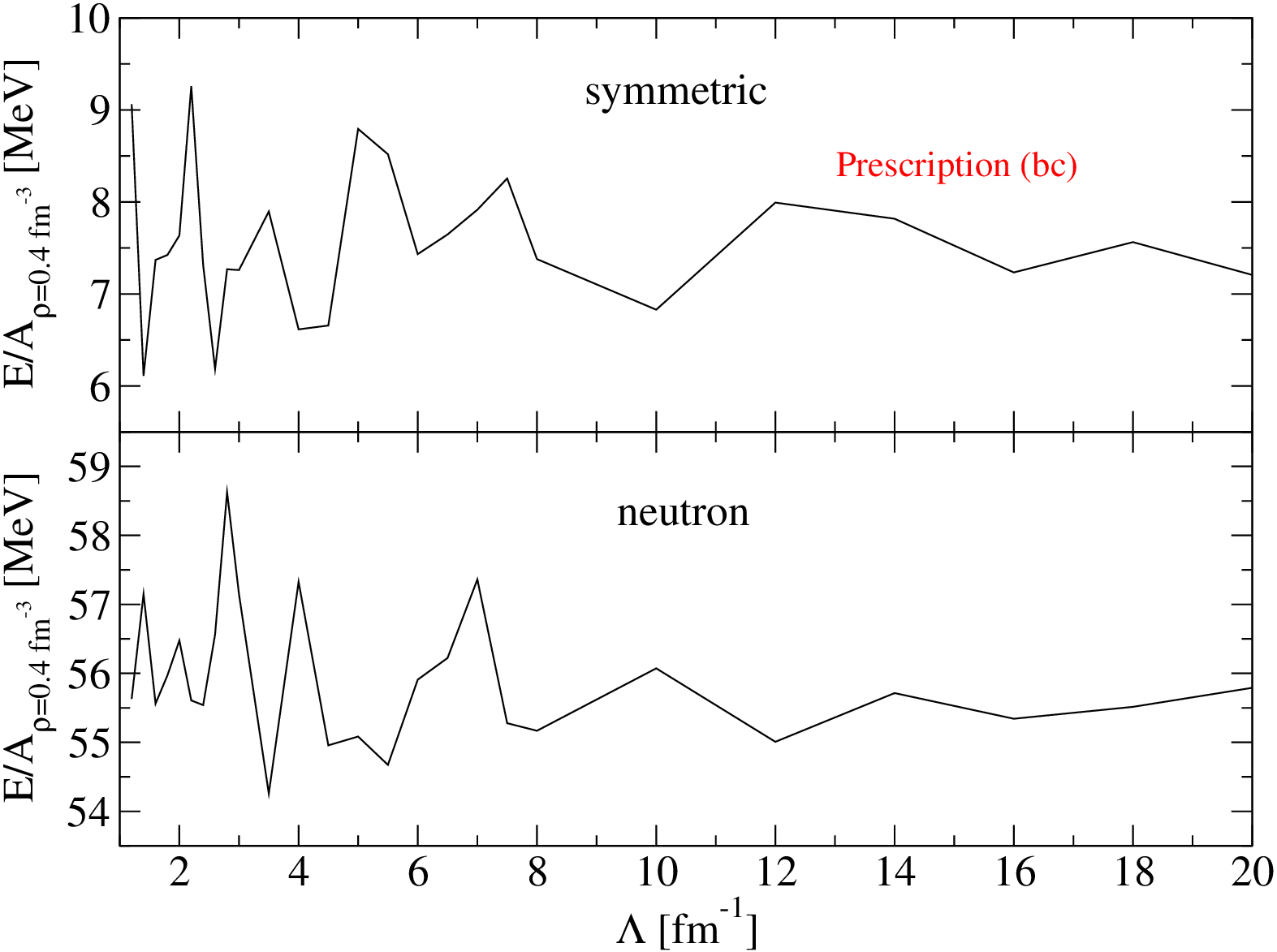}
\caption{Second--order EOSs obtained by using Eqs. (\protect\ref{bc}) and (\protect
\ref{bc2}) [scenario (bc)]  at the density value $\protect\rho =0.4$ fm$^{-3}$, as a function of the cutoff $%
\Lambda $. 
}
\label{plot_rg_bc}
\end{figure}

\begin{figure}[tbp]
\includegraphics[scale=0.35]{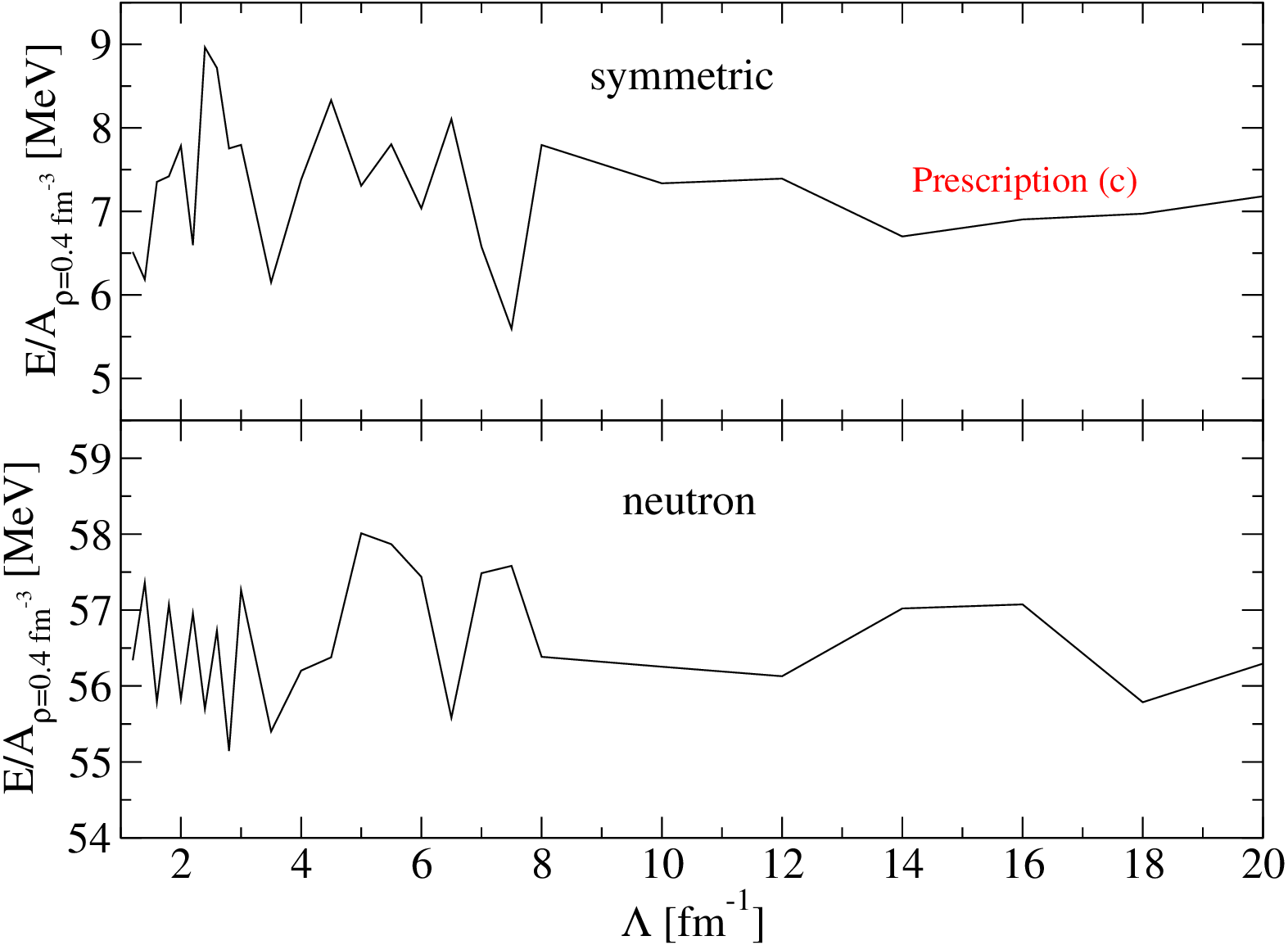}
\caption{Second--order EOSs obtained by using Eqs. (\protect\ref{a}) and (\protect
\ref{a2}) [scenario (c)] at the density value $\protect\rho =0.4$ fm$^{-3}$, as a function of the cutoff $%
\Lambda $. 
}
\label{plot_rg_c}
\end{figure}

\begin{figure}[tbp]
\includegraphics[scale=0.42]{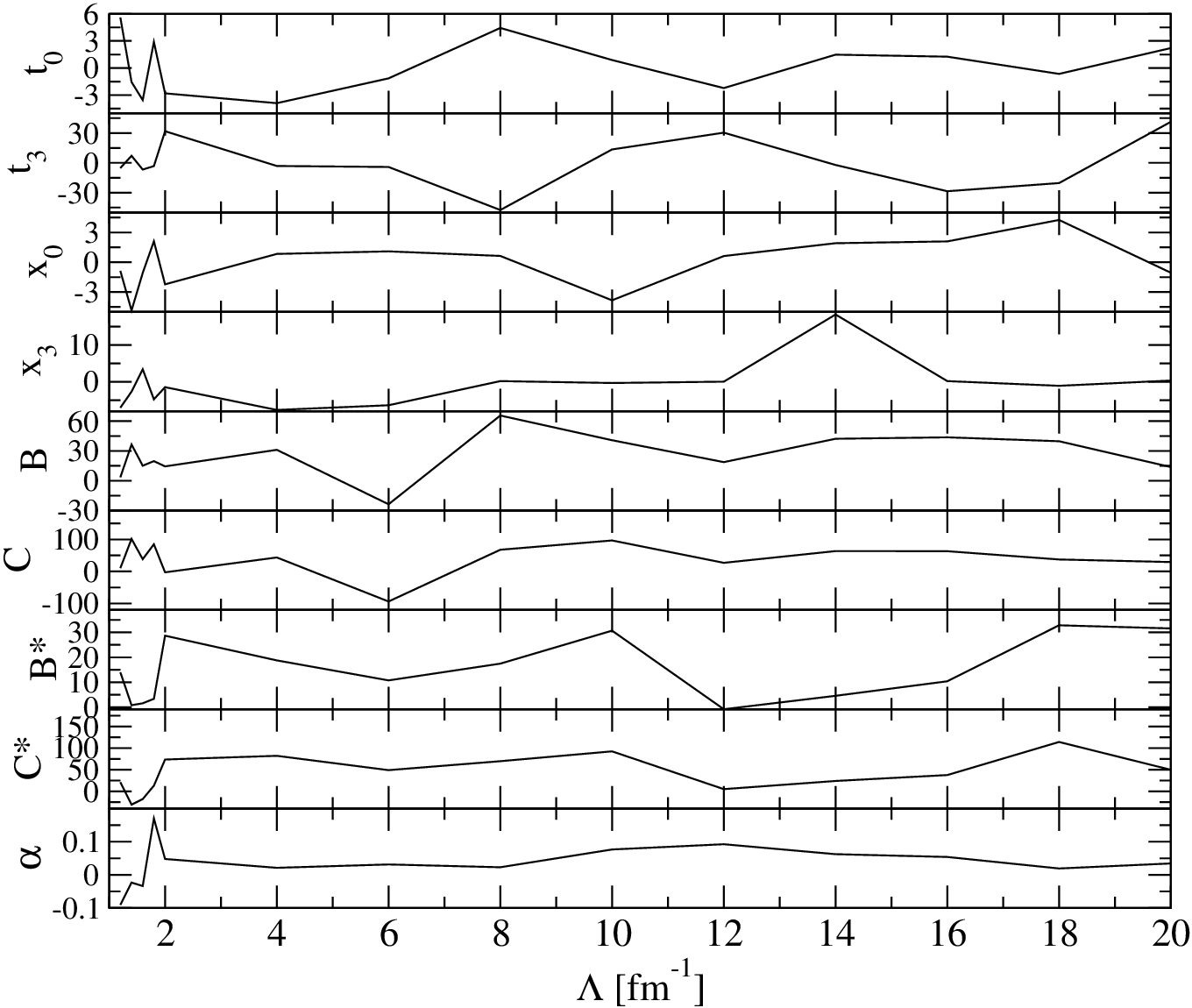}
\caption{Parameters in Eqs. (\protect\ref{bc}) and (\protect
\ref{bc2}) [scenario (bc)] as a function of the cutoff $\Lambda$. Here the units of parameters are those listed in Table \ref{tnlob}.}
\label{parabc}
\end{figure}

\begin{figure}[tbp]
\includegraphics[scale=0.35]{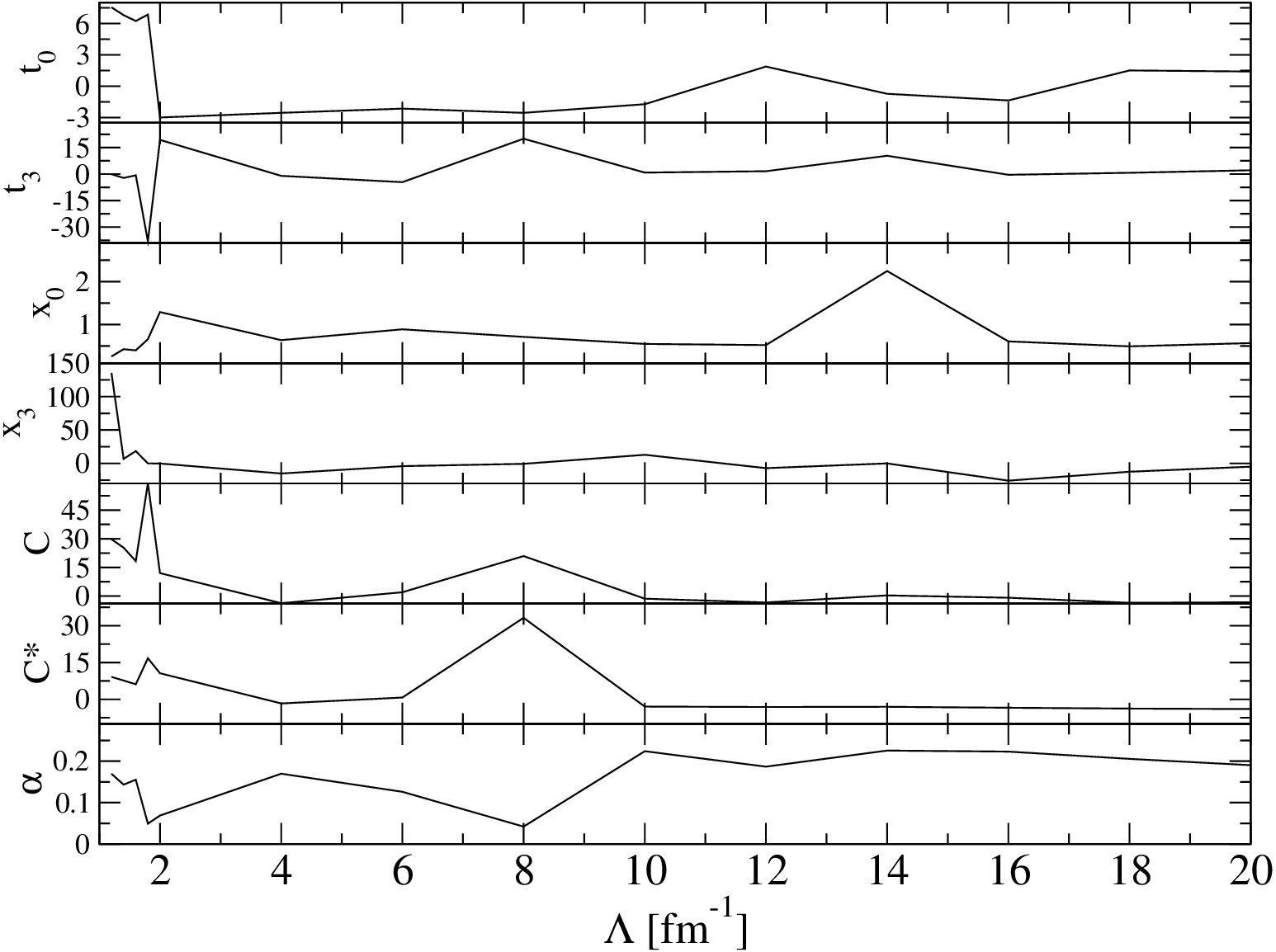}
\caption{Parameters in Eqs. (\protect\ref{a}) and (\protect
\ref{a2}) [scenario (c)] as a function of the cutoff $\Lambda$. Here the units of parameters are those listed in Table \ref{tnloc}.}
\label{parac}
\end{figure}

Although in the present work the interactions are treated perturbatively and the small difference between the LO and NLO EOSs suggests that the power counting should be straightforward, the fact that the LO interaction $V_{\rm LO}$ is not derived from an underlying microscopic theory and the presence of $V_{\rm NLO}$ counter terms leave the whole theory into the danger that what is generated could be nothing but just another phenomenologically fitted functional. Therefore, an EFT-based power counting analysis is necessary. For an explicit determination of the power counting, a RG-analysis needs to be performed first. Here, we performed a RG-analysis for the two scenarios where the cutoff dependence is still present. The cutoff dependence at the density $\rho =0.4$ fm$^{-3}$ is plotted
as a function of the cutoff $\Lambda $ in Figs. \ref{plot_rg_bc} and \ref%
{plot_rg_c}, where the EOSs are obtained by Eqs. (\ref{bc})-(\ref{bc2})
and Eqs. (\ref{a})-(\ref{a2}), respectively. In addition, the running of parameters is plotted as a function of cutoff $\Lambda$ in Fig. \ref{parabc} and Fig. \ref{parac} for scenario (bc) and (c), respectively.
 Note that the adjustment is performed up to $\rho=0.3$ fm$^{-3}$, so the results at $\rho=0.4$ fm$^{-3}$ are predictions. As one can see, the cutoff dependence is reduced at higher $\Lambda$ in both cases. In addition, a similar convergence pattern is observed. However, due to the uncertainty generated by the large number of parameters (nine for the case in Fig. \ref{plot_rg_bc} and seven for the case in Fig. \ref{plot_rg_c}), the convergence patterns in both cases are not quite smooth. This might give rise to potential problems in performing a full power counting analysis as introduced, for example, in Ref. \cite{gries}. Nevertheless, such an analysis is still of interest and should be performed at NLO and NNLO level to give further confirmation to our approach. We leave it as a future work.
Finally, for $\alpha \geq \frac{1}{6}$, one needs to consider also the mean--field
contributions coming from the $t_1$ and $t_2$ terms. They contribute at LO as $\frac{\theta _{s}}{4\pi
^{2}}k_{F}^{5}$ and $\frac{\theta _{s}-\theta_{v}}{4\pi
^{2}}k_{F}^{5}$ to the EOS of symmetric and pure neutron matter, respectively, where 
\begin{eqnarray}
\theta _{s}=\frac{1}{10}\left[ 3t_{1}+t_{2}(5+4x_{2})\right], \notag \\
 \theta _{v}=\frac{1}{10}\left[ t_{1}(2+x_1)+t_{2}(2+x_{2})\right].
\end{eqnarray}%
We have repeated the fit for $\alpha \geq \frac{1}{6}$ for the above three cases. However, we found
that, despite the presence of four additional parameters, the $\chi ^{2}$  increases for most of the cutoff values in the three
cases for $V_{\rm NLO}$. This suggests that the inclusion of the $t_{1}$, $t_{2}$ terms
should be deferred to NNLO.

A final point we wish to stress is that the sets of parameters listed in Tables \ref{tnloa}-\ref{tnloc} are obtained through a fit to the SLy5 EOS of symmetric and pure neutron matter with 10 points ranging from $\rho=0-0.3$ fm$^{-3}$. We have changed the number of points from 9 to 12 and found that the parameters are stable with respect to the number of fitting points. However, due to the large number of parameters and to the fact that the fits are performed only to two EOSs, there exist other sets of parameters which generate slightly ($<1\%$) larger $\chi^2$. Thus, we cannot guarantee that the parameters listed in Tables \ref{tnloa}-\ref{tnloc} are the final values to be used in all applications. A full determination of parameters is only possible with a general fit to both nuclear matter and finite nuclei, which we defer to a future work. Nevertheless, when another set of parameters (with slightly larger $\chi^2$) is adopted, we observed that the convergence pattern as listed in Figs. \ref{plot_rg_bc} and \ref{plot_rg_c} is unchanged, that is, the oscillation with respect to the cutoff $\Lambda$ becomes smaller at higher $\Lambda$. Also, after canceling the divergence by the contact terms, it could be possible to keep a subset of 
parameters cutoff invariant. For example, one could try to keep $t_3$, $x_3$, and $\alpha$ cutoff invariant 
in the scenario (bc) and $\alpha$ cutoff invariant in the scenario (c). Decreasing the number of parameters for the fit might indeed help to reduce the fluctuations seen in Figs. 9-12. This kind of test will be performed in a future work to gain more insight toward establishing an EFT-based functional.

\section{Conclusions}

We have proposed a new approach to generate an effective interaction up to NLO in
the EDF framework. Two tools from EFT, renormalizability
and RG-analysis, are utilized to construct and analyze the new effective interaction. Under the condition that the renormalizability is guaranteed, we explored three possible scenarios for the NLO counter terms. We found that all three scenarios produce second--order EOSs with similar quality, which indicates that our EOS up to NLO is independent of the regularization scheme.   
Benchmark symmetric and neutron matter EOSs can be reproduced in our approach within $\chi^2<5$ for a wide range of cutoffs. 

There are many possibilities to extend the current study. 
In particular, the extra parameters provided by the counter terms may be determined in a future work by a fit to properties of some selected finite nuclei. Also, a more conclusive power counting might be drawn after higher--order (e.g., NNLO) contributions are included, which will be addressed in a future work. As an interesting step, it is worth mentioning that the third--order perturbation terms associated to Skyrme forces have been derived recently in Ref. \cite{Kai17}. 

\begin{acknowledgments}
We thank U. van Kolck for useful discussions and suggestions. This research was supported 
by the European Union Research and Innovation program Horizon 2020 
under grant agreement no. 654002. 
\end{acknowledgments}


\appendix



\end{document}